# Identifying The Potential Biosphere of Mars

Eriita G. Jones and Charles H. Lineweaver

*Planetary Sciences Institute, Research School of Astronomy and Astrophysics, The Australian National University, Canberra, ACT, 0200, Australia*

**Summary:** Our current knowledge of life on Earth indicates a basic requirement for liquid water. The locations of present liquid water are therefore the logical sites to search for current life on Mars. We develop a picture of where on Mars the regions with the highest potential near-surface liquid water abundance can be found through a study of gullies. We also use rampart craters to sound the depth of water ice on Mars and where the highest concentrations of water ice occur. We estimate that low latitude gullies and rampart craters with depths > ~100 m at $|30^o|$ latitude, > ~1.3 km at $35^o$ and >~2.6 km at $40^o$ latitude will give access to current liquid water environments capable of supporting microbial life. Our data is most consistent with formation of these gullies through shallow aquifer discharge. These features should therefore be high priority targets for further study and high–resolution imaging with HiRISE.

## Introduction

**Looking for life**

Life on Earth is robust and inhabits a vast range of environments exposed to liquid water [1]. Microbial life has been found in many environments where the conditions are far beyond those hospitable to humans. Existence of terrestrial life in extremely cold environments and dark, dry environments up to 5.3 km below the surface have important implications for planetary exploration [2,3]. The nearest potentially habitable planet - Mars - has an average surface temperature ~70 degrees colder than Earth's average (~20 degrees colder than the average in Antarctica, the coldest region on Earth) and thus the majority of liquid water environments (if they exist) will lie beneath the Martian surface. Life is able to survive in temperatures far below the freezing point of pure liquid water due to the presence of salts (which depress the freezing point [4]) and the existence of thin films of liquid water in intergranular veins [5]. In deep intraterrestrial environments microbial life is found in liquid water environments within the pores of sediments and unconsolidated material as well as rock fractures and fluid inclusions. These adaptations of life to such hostile planetary environments are a powerful indicator of the habitability potential of Mars for terrestrial-like life.

NASA's MEPAG (Mars Exploration Program Analysis Group) report (*Mars Scientific Goals, Objectives, Investigations and Priorities: 2006* [6]) has identified sites with geologically recent water on Mars as being important astrobiological targets for future exploration. We have addressed the goal of identifying and describing environments potentially favorable to terrestrial-like life in this study. Gullies (see Figure 1) and rampart craters (see Figure 2) are ideal astrobiological targets as gullies are most likely associated with water and rampart craters indicate the presence or former presence of near-surface ice. Further, crater walls are not only the preferred location for gully formation[7,8], they also allow future missions access to subsurface environments.

**Gully formation**

Gullies are strongly suggestive of surface fluid runoff that has occurred within the current Martian obliquity cycle [9]. The tilt of Mars' axis undergoes chaotic variations which have major effects on the Martian climate as they modulate the sublimation of the polar caps and the atmospheric pressure [10]. Furthermore, the light-toned flows observed in two gullies within the last 10 years [11,12] indicate that there may be modern groundwater environments in some regions of Mars. This makes them strong candidates in the search for current liquid water environments and microbial habitats on Mars (see Figure 1).

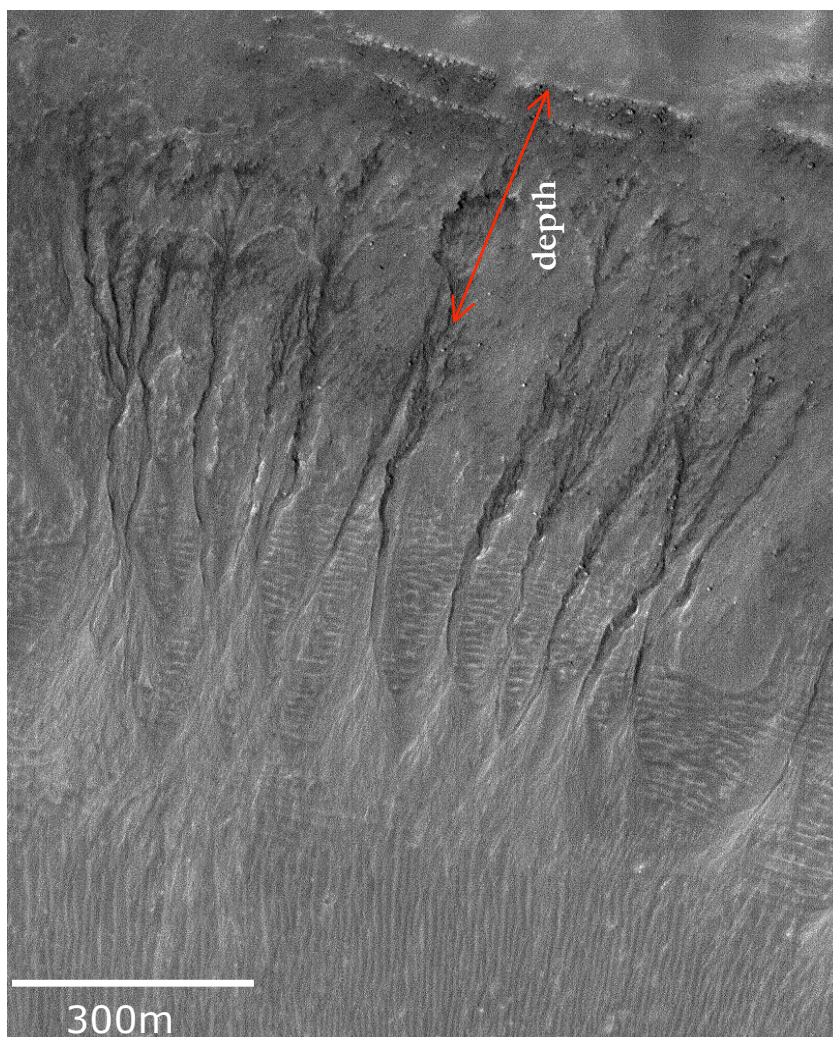

*Fig. 1: Martian gully system in crater in central Noachis Terra, located at 32.1°S, 12.9°W. Image shows that the alcoves of the gullies in this system are all at similar depths (~250m) beneath the overhanging surface. This may indicate the depth to groundwater in this region. The debris aprons of these gullies can be clearly seen. Image is M0C2-728; courtesy of NASA/JPL.*

**Rampart crater formation**

Rampart craters are an impact morphology that have lobate ejecta blankets elevated above the local terrain. The ejecta of rampart craters is strongly indicative of the deposition of material as a surface flow lubricated by fluid. Features such as flowlines, re-direction of the ejecta around obstacles, and distal thinning of the blanket with increasing radial distance from the central crater, all indicate that entrained fluid was present in the material. This source fluid is believed to be liquid water from subsurface ice melted by the impact heating. As rampart craters most likely indicate the presence of water ice, they are strong candidate locales for both current liquid water (which may be found at the base of the permafrost [38]) and sub-freezing water analogous to environments on Earth which harbor thin films of liquid water and associated microbial life (see Figure 2).

**Water on Mars**

Mars currently has a sizable water abundance: the presence of hydrogen on Mars - either as free water or hydrated minerals – has been detected by the neutron detectors carried as part of the gamma-ray spectrometer (GRS) and the Russian high-energy neutron detector (HEND) on board NASA's Mars Odyssey spacecraft. Neutrons are generated within the top 2 m of the Martian surface from the interaction of cosmic rays. The neutron flux is decreased through neutron absorption and retardation by hydrogen atoms in the subsurface (as these have a comparable mass) and so the observed flux provides a direct measure of the hydrogen content. These results have indicated that the Martian permafrost has an increasing water ice content pole wards of 50-60$^o$ latitude [13,14] with the concentration varying between 13-80% by weight. The water content in the lower latitudes is significantly depleted, varying between 1-10% by weight.

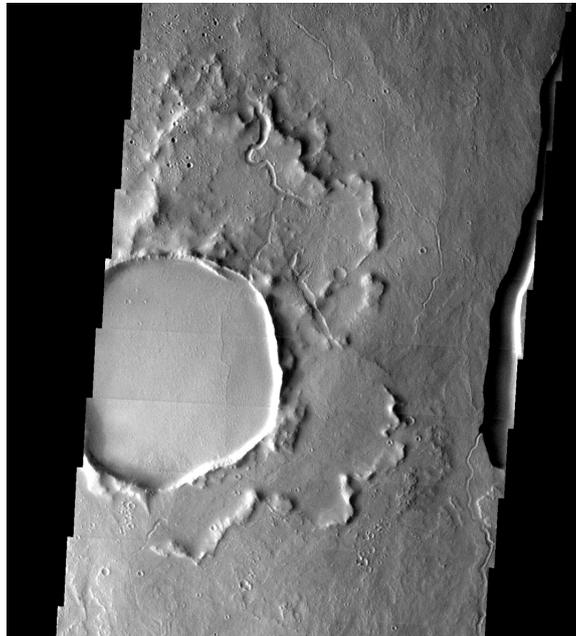

*Fig. 2: Image of rampart crater at latitude 16.8$^o$N, longitude 257.4$^o$E, northeast of Ascraeus Mons. The lobate ejecta blanket can be seen and its elevation above th elocal terrain in some regions is evident from the distal shadowing. In the upper part of the image a meandering channel can be seen carved into the ejecta, indicating the flow of water or potentially a lava fluid. Image taken by THEMIS onboard Mars Odyssey; ID: PIA04522; courtesy of NASA/JPL.*

In Mars' recent past (and to the present day) the pressure and temperature conditions on the surface have not been favourable for the stability of liquid water [11]. The Martian surface pressure is generally below the triple point except at the lowest elevations in the southern hemisphere (the Hellas and Argyre Basins) and the low latitude northern plains [39]. At any other locations liquid water on the surface would either freeze or sublimate. The surface temperature is generally too cold for liquid water (however daytime temperatures at the equator can climb above 0$^o$ C). Thus in order to look for stable reservoirs of liquid water and potential life we will need to excavate beneath the surface of Mars.

# Methodology

We investigated the global distribution of Martian gully depth with latitude. Our data is taken from surveys by [15,16,17,7,8]. Gully depth is the depth of the gully alcove bases beneath the local surface - the probable region where the gully fluid escaped onto the surface. We used gully depth measurements from these studies obtained by correlating elevation data from the Mars Orbital Laser Altimeter (MOLA) with images from the Mars Orbital Camera (MOC). The error in a single MOLA elevation is approximately 10 m (see http://ssed.gsfc.nasa.gov/tharsis/spec.html). As several footprints are used to create the elevation map for each gully system the combined vertical accuracy would be greater than 10 m. We binned the data into ten equal surface area latitude bins (six were occupied) and 100 m depth bins. The latter was deemed to be the limit at which we could be reasonably certain we had binned the data accurately from the published plots. Jennifer Heldmann (NASA Ames Research Centre) generously contributed digital versions of her northern and southern hemisphere data to this study.

We made a similar study of Martian rampart craters. The dataset for these craters was the "Catalogue of Large Martian Impact Craters"[1] compiled by Nadine G. Barlow. The catalogue is considered to be complete for craters with diameter greater than or equal to 5 km [18]. As we are interested in the signature of recent Martian water that may have persisted to the present day, the study focused on the youngest sample of craters that could be obtained from the catalogue. Craters that were determined as existing on terrain that pre-dates the Amazonian epoch were removed from the sample[2], however, some young (Amazonian aged) craters may have been discarded if they post-dated pre-Amazonian terrain. The Amazonian epoch (the present Martian epoch) has an estimated boundary age of 1.8-3.5 Ga [19] and thus all the ramparts within our sample are at most 3.5 Ga old. We estimated the rampart's excavation depth (d) - the maximum depth from which its ejecta blanket, and hence material believed to be volatile abundant, originated from – by this scaling of its diameter (D):

$$d = 0.12 D \qquad (1)$$

This value is an average of that used by [20] for simple and complex craters; and is at the top end of the range of relationships proposed by [21] for simple craters. We examined the dominant crater morphology (rampart or non–rampart) by the following dominance statistic:

$$\mu = \frac{N_{Craters} - N_{Ramparts}}{N_{Ramparts}} \qquad (2)$$

$N_{Craters}$ is the total number of impact craters and includes both ramparts and non-ramparts. The smaller the value of $\mu$ the stronger the rampart dominance (with a zero value indicating that all craters are ramparts). When $\mu > 1$ non - rampart craters are more frequent. Values of this test statistic were calculated over each of the three occupied latitude bands.

In addition, we have made a first estimate of the maximum range of conditions that could be present within the potential Martian biosphere (PMB). The term potential biosphere in reference to Mars means 'the portion of the planet that can potentially support life'. Our estimate of the PMB is described in terms of the range of pressure and temperature conditions that it encompasses and by the extent that it penetrates within the Martian regolith. This estimate of the Martian biosphere is based on our understanding of the environments on Earth that support life and the range of pressures and temperatures within these environments and so we investigated some extreme P-T conditions of inhabited environments on Earth. Different environments on Mars will have different potential for habitability, however, as the extent of the biosphere is dependent on both the physical and chemical properties of the subsurface and thus would vary both with latitude and locally. To introduce the depth dependence to our model an adoption of the thermal model published in [22] was utilised. This model calculates the subsurface temperatures from a combination of solar insolation and geothermal heating. Solar insolation is incorporated through the annual average surface temperature of the region and geothermal heating is quantified by the geothermal heat flow. Subsurface temperature and pressure are given by:

$$T(z) = T_s + zq/k \qquad (3)$$

$$P(z) = P_s + \rho g z \qquad (4)$$

where $z$ = depth (m); $T_s$ = annual mean surface temperature (K); $q$ = geothermal heat flow flow (= 30 mWm$^{-2}$); $k$ = thermal conductivity of soil (Wm$^{-1}$K$^{-1}$); $P_s$ = annual mean surface temperature (= 600 Pa); $g$ = Martian gravitational acceleration (= 3.73 ms$^{-2}$); and $\rho$ = density of regolith (kg m$^{-3}$). The values we used for these parameters are given in Table 1. Mellon and Phillips [22], however, incorporated only two 'extreme' soil compositions into their model, corresponding to highly desiccated and densely ice-cemented, respectively. A compositional gradient is more realistic. However, we used the model of [22] to determine the depth to 273 K at 30°, 60° and 90° latitude. We used decreasing average surface temperature with increasing latitude. Between 30-60° latitude we assumed a linear gradient in thermal conductivity, equivalent to assuming a linear gradient in soil composition with latitude. This allows us to estimate how the depth the 273 K isotherm varies with latitude, ignoring latitudinal variations. The thermal conductivities taken (see Table 1) are plausible representative values of the subsurface [7]. However, we acknowledge that a large variation of values would be expected. Thermal conductivities can be calculated from thermal inertia values measured by the Thermal Emission Spectrometer, valid for the top few cm of soil [41]. These values however give a range of 0.023 – 0.246 Wm$^{-1}$K$^{-1}$ which is smaller (by an order of magnitude) than the range of end-member models taken for the subsurface (see Table 1). This reflects the lack of constraints on subsurface soil composition and thus the estimates of the depth to 273K are only a rough guide of the plausible latitudinal behaviour.

---

[1]   This catalogue is available at ftp://ftpflag.wr.usgs.gov/dist/pigpen/mars/crater_consortium/ mars_craters_geographic/craters/all_catalogs/. The database file utilised was: Barlow_craters_all.dbf. Ramparts were identified as those with Rc (indicating lobate ejecta) or RF (indicating Rc with fractures cutting across the ejecta) in Column 7 of the database file.

*Table 1: Subsurface model parameters**

| Latitude | $T_s$ (K)[**] | $k$ (Wm$^{-1}$K$^{-1}$) | $\rho$ (kgm$^{-3}$) |
|---|---|---|---|
| 30° | 210 | 0.045[a] | 1650[c] |
| 60° | 180 | 2.4[b] | 2018[d] |
| 90° | 150 | 2.4 | 2018 |

[*]Data taken from [22] unless otherwise indicated.
[**]Taken from [23].
[a] Corresponds to dry particulate soil.
[b] Corresponds to densely ice-cemented soil.
[c] Corresponds to dry unconsolidated soil as found by the Viking Landers.
[d] Corresponds to ice-cemented soil.

# Results

**Shallow and abundant water**

Figure 3 reveals that on average gullies in the high latitudes were found to occur roughly 54% shallower (between ~90-190 m depth) than gullies equatorward of 53° (this trend was found by [17,7,15,16] in sub-samples of this study). Figure 4 reveals that gullies in both hemispheres are also less constrained (indicated by the 68% confidence interval) poleward of 53° than they are equatorward. This preference for gullies to form deeper in the warmer, near–equatorial latitudes strongly suggests a thermal control on their formation, in agreement with studies by [7,17,15,16]. The trend of decreasing average depth with increasing latitude in the northern hemisphere is in agreement with [17], however this trend is poorly constrained as the depth range for two of the latitude bins (37-53° and 53-90°) significantly overlap. Gilmore and Phillips [15] found the average depth of their sample between 26-38° to be within 200-400 m, in agreement with our results. In both hemispheres the average gully depth across all latitudes in this study was found to be approximately 300 m. In the northern hemisphere, the vast majority of alcove bases lie within 500 m of the upper reaches of the encompassing slope, in agreement with [8]. Any explanation of gully formation will have to be consistent with both the trend of maximum gully depth with latitude and the occurrence of gullies close to the surface in all latitudes. Furthermore no gullies in the sample occurred deeper than 2 km and the majority occur shallower than 800 m, indicating that they can only form in a restricted depth of regolith. This is despite the presence of many crater slopes extending to beyond 2 km depth (see Figure 4).

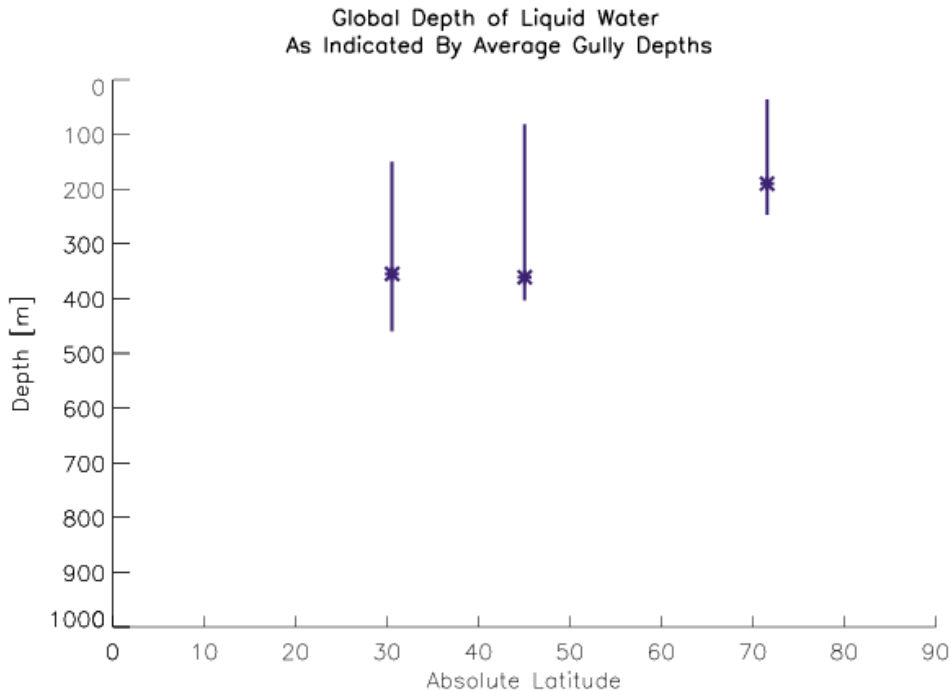

*Fig. 3: Average depth of 549 gullies in the three occupied latitude ranges. The vertical lines give a 68% confidence interval. The average depth of gullies is shallowest in the high latitudes.*

Figure 4 reveals that the average depth of young ramparts may decrease with increasing latitude (agreeing with the trend found by [24,25]). However, the significant overlap of the average depth ranges and the similar spread within each latitude range suggests that this trend is not significant. Most gullies form significantly deeper than the depth of the thermally active layer which seems inconsistent with formation by fluid generated from the melting of water-rich permafrost [42], or snow deposited during high obliquity [43]. Such meltwater could percolate to depths below the active layer but it is unlikely that it could remain liquid and form gullies at depths exceeding 100 m below where the fluid was generated. Gullies in the mid northern latitudes and high latitudes preferentially form at shallow depths (considerably shallower than the estimated average depth to 273K) and so may be most consistent with formation from shallow meltwater.

Observations of Martian craters reveal that the radial ejecta morphologies dominate in the very small (D<5 km) and very large (D>50 km) diameter ranges with lobate structures dominating in between [26]. Converting to excavation depth via Equation 1, we would then expect to see rampart craters dominating in the 600-6000 m excavation depth range. We have quantified the regions where Amazonian rampart craters have been the preferential impact crater morphology in Figure 5 using Equation 2. Figure 5 reveals that non-ramparts are the preferential impact morphology in the mid and low latitudes at the majority of depths, however sporadic concentrations of volatiles are seen at the depths where ramparts are more abundant. In the high latitudes rampart craters are more abundant than non-ramparts at all depths and the presence of an abundant volatile layer extending between 500-2200 m depth is indicated. Volatiles become gradually less abundant at greater depths as non-ramparts become more frequent. These results are in disagreement with that previously reported as the diameter ranges where ramparts dominate are found to be strongly latitude dependent and consistently much smaller than the 5-50 km range previously reported.

The results in Figure 5 strongly indicate the presence of a thick present day subsurface ice or fluid layer occurring polewards of 53°. The presence of this layer is indicated both by the dominance of ramparts at excavation depths down to approximately 2 km in these latitudes, and by the small values of $\mu$ (predominantly $\mu>0.5$, indicating 50% more ramparts than non-ramparts within these regions) which suggest that these results are significant.

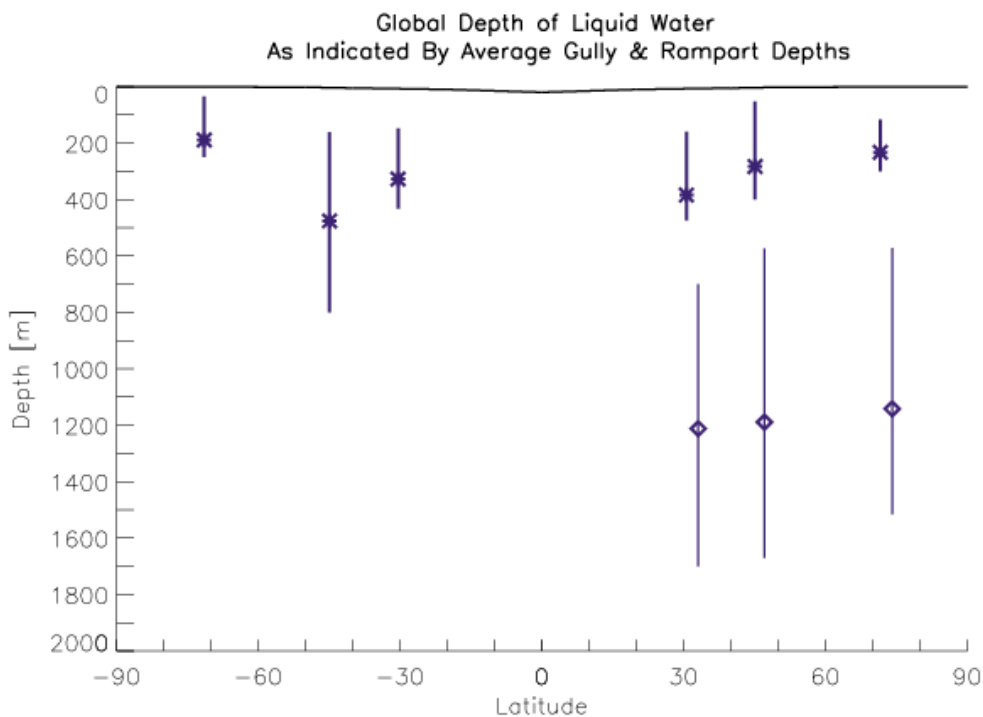

*Fig. 4: Average depth of 549 gullies (asterisks) and 1329 ramparts (diamonds) compared to the depth of the seasonally active layer (explained in text; derived using values given in [27] and an interpolation for equatorial latitudes) given by the black line . Vertical lines give a 68% confidence interval. Gullies are shallower in the higher latitudes, whereas rampart depths show little variation with latitude. On average ramparts are considerably deeper than*

*gullies and form well within the cryolithosphere (the Martian permafrost, i.e. the lithosphere that has a temperature below 273 K).*

Isolated regions of rampart crater dominance occur in the low- and mid–latitudes but their restricted depth suggests that they are not part of an extensive sub-surface volatile layer. Rather they may be indicative of depths where unusual geological conditions, perhaps the presence of an impermeable rock layer that prevented diffusion of volatiles into the atmosphere or deeper into the regolith, allowed ice or fluid to be stable for extended periods of time.

The average depths of gullies and ramparts are compared to the estimated depth of the seasonally active layer in Figure 4. This layer is the region of the regolith that is exposed to (seasonal) thermal cycling – freezing during the winter and thawing during the warmer months. No gullies occur within this region. However, this does not preclude their formation by meltwater that was generated within this layer and then percolated through the regolith to form the gullies at greater depths [42]. Near-surface ice is not abundant in the low latitudes [27] and so the melting of water-rich permafrost (by geothermal heating) would not contribute a significant amount of ground water. Furthermore the depth of most of these gullies exceeds the estimated minimal depth of a subsurface water aquifer in these latitudes [22] (see also Figure 7). It seems probable that the source fluid for the gullies originated predominantly from a shallow liquid water aquifer. As the upper layers of the regolith are in sub-freezing conditions, liquid water is most likely associated with water ice. Amazonian rampart craters (see Figure 5) indicated that the highest low latitude concentrations of water ice occur in several narrow ranges, encompassing both shallow (eg. 900m) and greater depths (eg. 1900m). Thus a subsurface aquifer may be highly localised but may promote the formation of both shallow and deep gullies.

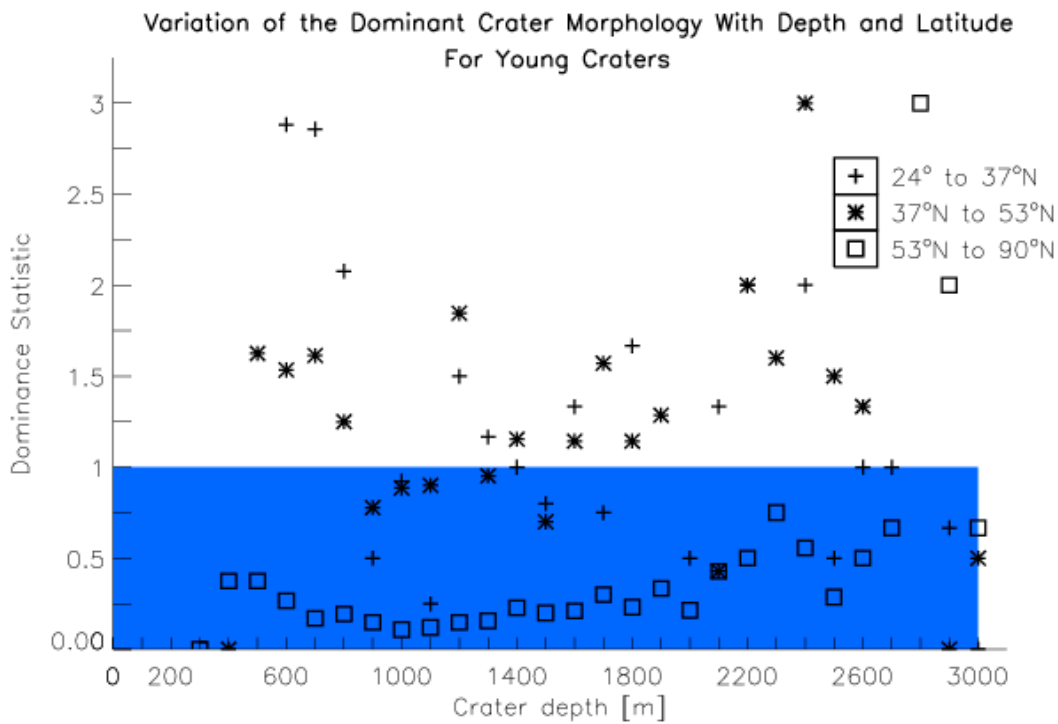

*Fig. 5: Variation of the dominant crater morphology with depth and latitude for our sample of young (Amazonian aged) craters. The blue region encompasses the range of µ values that indicate ramparts are more abundant than non-ramparts.*

In the mid- and high-latitudes gullies may have formed within the cryolithosphere, as estimates of the average surface temperatures and the thermal conductivity for ice-cemented soil place the 273K isotherm at depths beneath the 2km range over which gullies occur. Thus the regolith at these gully alcove bases may be in subfreezing conditions if the estimated thermal conductivity of [22] for ice-cemented soil is reasonable. The source fluid for shallow gullies is most likely meltwater from surface ice (with a high water concentration [27]) generated during warmer surface temperature. Shallow (on the order of tens of centimetres; see Figure 4) cracks in the permafrost related to seasonal freeze-thaw cycles in the rock strata may propagate to greater depths as the subsurface within the cracks is exposed, allowing an available passage for a rapid meltwater discharge. It seems unlikely, however, that this mechanism could form gullies several hundred metres below the subsurface in these latitudes. These deep gullies may originate from meltwater brought up from greater depths below the permafrost. This requires the presence of significant hydrostatic pressures to bring the liquid water to the surface. Finally young ramparts form within the cryolithosphere at all latitudes, consistent with their formation requiring subsurface water ice.

**Potential Martian biosphere (PMB)**

We examined the range of pressure and temperature conditions in environments which host terrestrial life to quantify the PMB (see Figure 6).

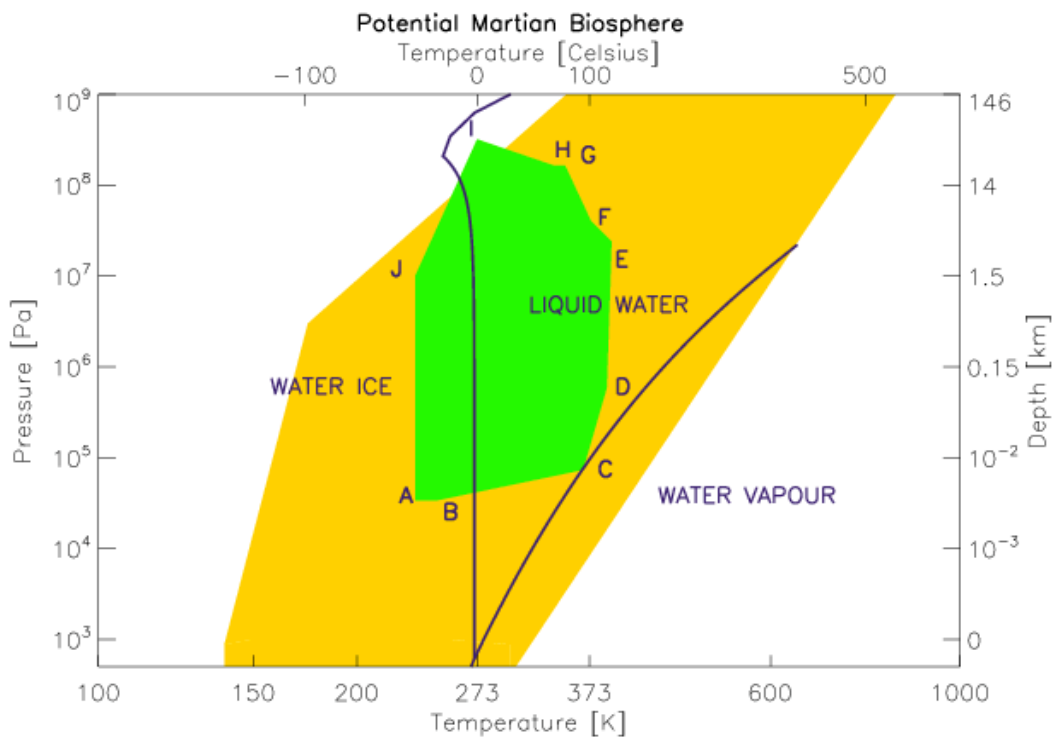

*Fig. 6: This figure shows our estimate of the Potential Martian Biosphere. The green region encompasses the environments known to be hospitable to terrestrial life. The letters correspond to the key in Table 2. The yellow region is an estimate of the area through which the Martian geotherms (functions giving the variation of temperature with depth/pressure) are expected to pass. As better estimates of these geotherms are obtained this region will be refined. The phase diagram of water is overlaid. The PMB may be accessed at around ~10 m in the warmer latitudes.*

The maximum observed temperature for terrestrial life is 121°C [28]. Our study found no convincing pressure limit as the high limiting pressures of terrestrial environments are due to either our limits in subsurface excavation within the crust or to the oceanic crust boundary.

Over time as more extremophiles are discovered this model can be updated to give necessary justification to NASA's strategy of 'follow the water'. Alternatively, if life is found to be pressure or temperature limited within liquid water environments, this model will essentially provide a map for the future search for extant life.

Figure 6 indicates that many of the conditions hospitable to terrestrial life may be reached in the Martian crust. Using our adaptation of the subsurface pressure - temperature model by [22] the range of estimated depths at which liquid water is stable between ~20–40° latitude is 90-2600 m (see Figure 7). Within the limits of our depth resolution (100 m) all gullies occurred within this range of depths. Above ~50° latitude, however, the lower average surface temperatures and the presence of low concentrations of water ice in the shallow subsurface may combine to lower the 273K isotherm to depths greater than the range at which gullies occur (as indicated in Figure 7). At temperatures below 121 °C life is found to inhabit the full range of conditions within the region of liquid water stability. Thus low latitude gullies may indicate current liquid water environments that are potentially favourable to microbial life and so should be considered high priority astrobiological targets. Several low latitude gullies within our sample are shown in recent Mars Orbital Camera (MOC) imagery to be associated with features indicative of either present or recent surface ice (indicating potential replenishment of the source aquifer) or extensive (possibly ancient) fluvial activity in the region. We suggest that gully systems which have alcove base depths below the estimated depth to liquid water, for example in MOC images m1002624, m0903654, m1800182, m0204149 and m1400457, should be investigated further with high resolution imagery to determine their viability (in terms of slope angle and potential rover hazards such as boulders) as candidates for future rover exploration.

*Table 2: Key for Figure 6*

| Label | Description | Temp. (K) | Press. (Pa) | Ref. |
|---|---|---|---|---|
| A | Mt Everest | 233 | $3.37 \times 10^{-4}$ | [29,30] |
| B | Mt Everest | 246.89 | $3.37 \times 10^{-4}$ | [29,30] |
| C | Hydrothermal vents in Octopus Spring, Yellowstone National Park | 356 | $7.33 \times 10^{-4}$ | [31,32] |
| D | Kuieshan thermal vents | 389 | $5.88 \times 10^{6}$ | [33] |
| E | Maximum known temperature for life. Strain121, Finn vent, Mothra hydrothermal vent field (2270m depth), Juan de Fuca Ridge | 394 | $2.39 \times 10^{7}$ | [28,34,31] |
| F | Bacterial isolates | 373 | $4.00 \times 10^{7}$ | [35] |
| G | Bacteria within 5278m deep borehole | 348 | $1.64 \times 10^{8}$ | [3,36] |
| H | Bacterial within 5278m borehole | 338 | $1.64 \times 10^{8}$ | [3,36] |
| I | Marianas Trench (deepest ocean trench) | 275 | $3.20 \times 10^{8}$ | [37] |
| J | Minimum known temperature for life. Microbes in 1249m deep borehole in Vostok ice. | 233 | $10^{7}$ | [2,4,36] |

Nearly all young ramparts (~90 %) in the low latitudes excavate within the stability region of liquid water as they are deeper than 600 m. At higher latitudes nearly all ramparts excavate into (water) permafrost as they are shallower than the depth to the 273 K isotherm in Figure 7. Thus the slopes of low latitude ramparts give access to potential current liquid water environments and conditions favourable to life, particularly within the depth ranges given above (where ramparts were the dominant crater morphology). Higher latitude ramparts still have important astrobiological implications, however, as terrestrial life can be found in environments where the freezing point of water has been depressed [2].

## Conclusions

These results allow us to draw some conclusions on the possible locations of the shallowest abundant water on Mars. The abundance of gullies in the high polar latitudes and southern polar pits suggests a reasonable water abundance here within the past million years. These gullies form at shallow depths in subfreezing conditions and so may form through seasonal snow/ice allowing them to be frequently exposed to liquid water. When active these gullies may provide the most accessible source of liquid water and if they are repeatedly re-activated they may provide a hospitable environment to psychrophyllic bacteria. Gullies in the low latitudes of both hemispheres form at depths close to the 273 K isotherm [22] and at least two have been recently re-activated [12].

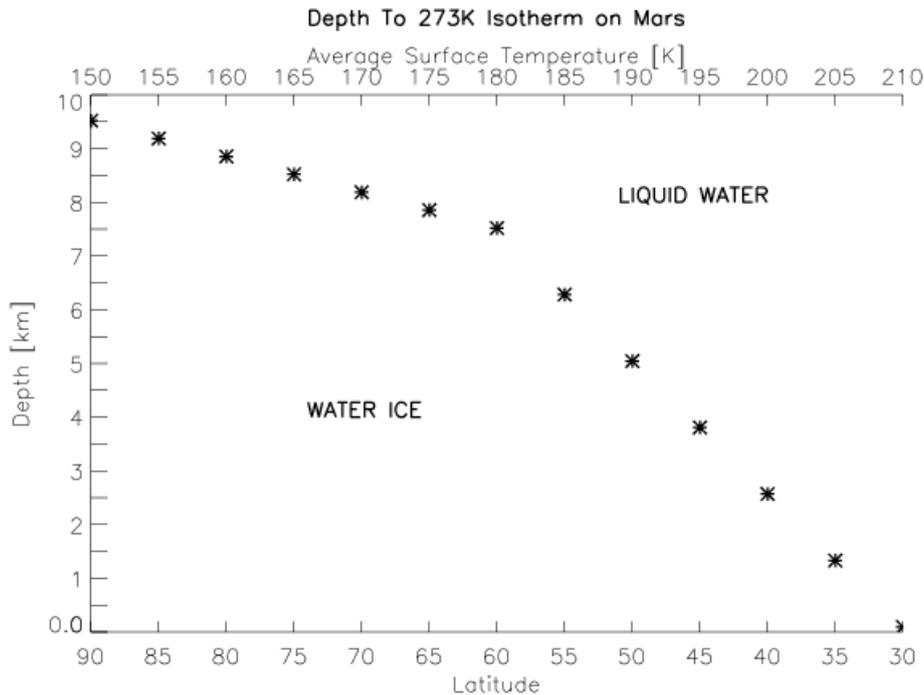

*Fig.7: Estimated depth to the 273 K isotherm. This estimate assumes that soil is dry and desiccated near the equator to densely ice-cemented at the poles. A discontinuity occurs at ~60º latitude [27] due to the stability of near - surface ground ice.*

As these gullies most likely form through a liquid water aquifer near the alcove base depths they may provide the most accessible stable source of liquid water and terrestrial – like environment on Mars. The walls of rampart craters in the low latitudes allow access to depths where the subsurface temperature is predicted to be above 0ºC, however, their distribution in these latitudes suggests that subsurface volatiles are highly localised. In the high latitudes young ramparts indicate areas of recent and abundant water ice (and potential liquid water environments at the base of the cryolithosphere). Some gully systems were suggested for further high resolution investigation to determine their viability as candidates for future rover exploration.

Acknowledgment: We thank three reviewers for extensive comments that improved the paper.